\documentclass[pra,showpacs]{revtex4}

\def\pa{\partial}
\def\bra{\langle}
\def\ket{\rangle}

\begin{document}

\title{Quantum metastability in a class of moving potentials}
\author{Chung-Chieh Lee and Choon-Lin Ho}
\affiliation{Department of Physics, Tamkang University, Tamsui
25137, Taiwan}
\date{Aug 15, 2001}

\begin{abstract}
In this paper we consider quantum metastability in a class of
moving potentials introduced by Berry and Klein. Potential in this
class has its height and width scaled in a specific way so that it
can be transformed into a stationary one.  In deriving the
non-decay probability of the system, we argue that the appropriate
technique to use is the less known method of scattering states.
This method is illustrated through two examples, namely, a moving
delta-potential and a moving barrier potential. For expanding
potentials, one finds that a small but finite non-decay
probability persists at large times. Generalization to scaling
potentials of arbitrary shape is briefly indicated.
\end{abstract}
\maketitle

\section{Introduction}

An interesting issue in cosmology is the evolution of metastable
states in the early universe according to the original version and
its variants of the inflationary models \cite{Guth,La}. In these
models inflation of the early universe is governed by a Higgs
field trapped in a metastable state.  Inflation ends when the
metastable state decays to the true ground state of the universe.
During inflation the universe expands exponentially.  It is thus
obvious that the metastable state of the Higgs field is trapped in
a rapidly varying potential.  The problem is therefore a truly
time-dependent one. However, owing to the inherent difficulties of
the problem, more often than not one considers the decay of the
Higgs field in a quasi-stationary approximation, in which the
decay is studied assuming a static potential \cite{Kolb}. Surely
this approximation is hard to justify, but for the present one has
to be content with it. Ultimately one hopes to be able to tackle
the non-stationary case. To this end, it is desirable to gain some
insights first by studying metastability in time-dependent
potential in simple quantum-mechanical models.

Roughly speaking, time-dependent potentials can be divided into
three classes.  In the first class we have potentials with
time-dependent strength. When the strength is small, the
Schr\"odinger equation can be solved by time-dependent
perturbation theory. Almost all textbook examples belong to this
type. When the strength of the potential is not small, other
methods of solution must be sought. For example, solutions of
time-dependent harmonic oscillator \cite{LR} and time-dependent
linear potential \cite{Guedes} can be obtained by the method of
invariant. We note here that the interesting phenomenon of quantum
tunneling induced by an externally driven field has also been
examined experimentally and theoretically \cite{Keay,Landauer,Wa}.
The second class of potentials involves time-dependent boundaries.
Unlike the first class, this class of potentials attracts much
less attention, and almost all previous works in this area
concerned only the simplest of all cases, namely, an infinite
potential well with a moving wall \cite{Makowski,Dodonov}.  The
last class is the combination of the previous two classes.

We believe that the barrier potential in an inflationary universe
is non-stationary, not only the barrier height but also the
barrier width should be changing as time elapses. However, it will
be extremely difficult to study metasbility in such a
time-dependent potential in full generality. Thus it would be
helpful if the quantum tunneling effect could be studied in any
class of moving potential, special though it is, as a step to
understanding the decay of a non-stationary metastable system.

In this paper we consider quantum metastability in a class of
scaling potentials which allows one to apply techniques used in
the corresponding problem with stationary potentials.  This class
of potentials was introduced by Berry and Klein \cite{Berry}.
Potentials in this class have their heights and widths scaled in a
specific way so that one can transform the potential into a
stationary one.

The organization of the paper is as follows. In Sec.~II. we give a
general discussion of the solutions of the Schr\"odinger equation
with the scaling form of time-dependent potential introduced in
\cite{Berry}.  It is argued that the most suitable technique for
studying quantum metastability in such kind of potential is the
less known method of scattering states. Two simple examples of
such metastable systems, a moving delta-potential and a moving
square barrier, are investigated in Sect.~III and IV,
respectively.  Generalization to arbitrary barrier is briefly
discussed in Sec.~V. Sec.~IV. concludes the paper.

\medskip

\section{Schr\"{o}dinger equation with a scaling potential}

We shall consider the problem of quantum metastability of a
particle of mass $m$ trapped in a moving potential $V(x,t)$.  We
assume that the potential $V(x,t)$ is of the scaling form proposed
by Berry and Klein \cite{Berry}, namely, $V(x,t)={\bar
V}(x/L(t))/L^2(t)$, where $L(t)$ is a time-dependent scaling
factor.  The Schr\"odinger equation is
\begin{eqnarray}
i\hbar \frac{\pa\Psi(x,t)}{\pa t}
= \left[ -\frac{\hbar^2}{2m}\frac{\pa^2}{\pa x^2}
 + \frac{1}{L^2(t)}\bar{V}\left(\frac{x}{L(t)}\right)\right]
 \Psi(x,t)~.
\label{eqs-1}
\end{eqnarray}
So far solution of Eq.(\ref{eqs-1}) is restricted mostly to the
special case in which $\bar V$ has the functional form of an
infinite potential well, i.e. $V(x,t)$ is an infinite well with a
moving wall \cite{Makowski,Dodonov}.  In this case the scaling factor $L^2(t)$
in front of $\bar V$ is immaterial.

For our purpose, we shall assume $\bar V$ to have the generic
shape of a potential well which is impenetrable to the left and
has a finite barrier to the right, much like that usually employed
in the discussion of alpha-decay.  We further assume $L(t)$ to be
a linear function of time:
\begin{eqnarray}
 L(t) = L_{0}+vt~,\ v={\rm constant}.
\label{eqs-2}
\end{eqnarray}
Of course, for $v<0$, the problem is meaningful only for time
duration $0<t<L_0/|v|$.  Eq.(\ref{eqs-1}) cannot be solved by
separating the time and spatial coordinates. However, for the
scaling form of $V(x,t)$ in Eq.(\ref{eqs-1}) and the linear form
of $L(t)$, separation of variables can be achieved through a
series of transformations introduced in \cite{Munier,Berry} (see
also:\cite{Makowski,Dodonov}). One first transforms the coordinate
frame into a rescaled frame with a rescaled coordinate $\bar{x}$
defined by
\begin{eqnarray}
\bar{x}(t) \equiv \frac{x}{L(t)}\ .
\end{eqnarray}
In this frame the Schr\"odinger equation becomes
\begin{eqnarray}
 i\hbar\frac{\pa}{\pa t}\Psi(\bar{x},t) = \left[
 -\frac{\hbar^2}{2mL^2}\frac{\pa^2}{\pa x^2}
 + i\hbar \frac{v}{L}\bar{x}\frac{\pa}{\pa \bar{x}}
 + \bar{V}(\bar{x}) \right] \Psi(\bar{x},t)\ .
\label{eqs-4}
\end{eqnarray}
Eq. (\ref{eqs-4}) can be further simplified by the following
transformation
\begin{eqnarray}
\Psi(\bar{x},t) = \frac{1}{\sqrt{L(t)}}\
 e^{\frac{im}{2\hbar}Lv\bar{x}^2}
 \Phi(\bar{x},t)\ ,
\label{eqs-5}
\end{eqnarray}
and the introduction of a new time variable $\tau$
\begin{eqnarray}
 \tau = \int_{0}^{t}\frac{ds}{L^2(s)} = \frac{t}{L_{0}L(t)}\ .
\label{eqs-6}
\end{eqnarray}
After substituting Eq.(\ref{eqs-5}) and (\ref{eqs-6}) into
(\ref{eqs-4}), we obtain the equation
\begin{eqnarray}
 i\hbar\frac{\pa}{\pa \tau}\Phi(\bar{x},\tau)
 = -\frac{\hbar^2}{2m}\frac{\pa^2}{\pa \bar{x}^2}\Phi(\bar{x},\tau)
 +\bar{V}(\bar{x}) \Phi(\bar{x},\tau)\ ,
\label{eqs-7}
\end{eqnarray}
which resembles the Schr\"odinger equation with a stationary
potential. Eq.(\ref{eqs-7}) can be solved by separation of
variables
\begin{eqnarray}
 \Phi(\bar{x},\tau) = \Phi(\bar{x})e^{-\frac{i}{\hbar}\bar{E}\tau}\ ,
\label{eqs-8}
\end{eqnarray}
where $\Phi(\bar{x})$ satisfies the eigenvalue equation
\begin{eqnarray}
 \left[
 -\frac{\hbar^2}{2m}\frac{d^2}{d x^2}
 + \bar{V}(\bar{x}) \right] \Phi_{k}(\bar{x})
 = \bar{E}_{k}\Phi_{k}(\bar{x})\ .
\label{eqs-9}
\end{eqnarray}
Once Eq.(\ref{eqs-9}) is solved exactly in the rescaled frame, the
exact wave function in the original frame is then given by
\begin{eqnarray}
 \Psi_{k}(x,t) = \frac{1}{\sqrt{L(t)}}\
 e^{\frac{im}{2\hbar}\frac{v}{L}x^2}\
 e^{-\frac{i}{\hbar}\frac{1}{L_{0}L}\bar{E}_{k}t}\
 \Phi_{k}\left(\frac{x}{L}\right)\ .
\label{eqs-10}
\end{eqnarray}
The set of solutions (\ref{eqs-10}) is complete and orthonormal
\begin{eqnarray}
 \bra \Psi_{k}(x,t)|\Psi_{l}(x,t) \ket
 = \bra \Phi_{k}(\bar{x})|\Phi_{l}(\bar{x}) \ket
 = \delta_{kl}\ ,
\end{eqnarray}
so using this set of solutions we can find a solution satisfying
any initial condition.  Furthermore, if an initial state $\Psi
(x,0)$ is expressible in the basis $\{\Psi_k\}$ as
\begin{eqnarray}
\Psi (x,0)=\sum_k c_k \Psi_k (x,0)~, \quad\quad c_k=\bra
\Psi_k(x,0)|\Psi (x,0)\ket~, \label{comb1}
\end{eqnarray}
then at a later time $t$ the state is
\begin{eqnarray}
\Psi (x,t)=\sum_k c_k \Psi_k (x,t)~. \label{comb2}
\end{eqnarray}

We have now succeeded in transforming the original time-dependent
Schr\"odinger equation into a time-independent one.  The problem
of calculating the decay probability of a particle confined in
$V(x,t)$ at time $t$ is reduced to the corresponding problem with
a static potential ${\bar V}(\bar{x})$.  Hence techniques used in
the time-independent potential for calculating decay rate can be
borrowed.

However, there are some subtleties.  Naively, one is tempted to
employ the most well-known method, namely, the complex eigenvalue
method, proposed by Gamow in his studies of the alpha-decay
\cite{Gamow}. In this approach an ``outgoing wave boundary
condition" is imposed on the solutions of the Schr\"odinger
equation for the particle trapped in the well.  That means
incoming plane wave solutions outside the potential well are
discarded right from the beginning. This procedure naturally leads
to an eigenvalue problem with complex energy eigenvalues.  One
then relates the imaginary parts of the energy to the decay rate.
While the complex eigenvalue method is straightforward and
physically reasonable, it suffers from some conceptual
difficulties \cite{Holstein}. For example, how can energy
eigenvalues be complex as we are dealing with a Hermitian
Hamiltonian?  Also, the eigenfunctions are not normalizable, a
difficulty directly related to the eigenvalues being complex.
Furthermore, the particle trapped in the well cannot be in an
eigenstate of the system in the first place, since such states are
not completely confined at $t=0$.

Apart from the difficulties mentioned above, the complex
eigenvalue method cannot be employed in our problem for other
reasons. First, the problem we are interested in is an
intrinsically time-dependent one, with a non-conservative
Hamiltonian.  Hence energy eigenvalues and eigenstates lose their
meanings altogether ($\bar E$ in Eq.(\ref{eqs-9}) is not an energy
eigenvalue). Second, the ``outgoing wave boundary condition",
essential to Gamow's method, cannot be imposed in our case.  The
reason is as follows. As we discussed before, in order to fix a
moving potential we need to transform our problem to a
corresponding static one in a rescaled frame.  But in this frame
the meaning of incoming or outgoing plane wave is rather obscure.
In fact, it can be checked that an outgoing plane wave in the
original $x$-$t$ (rescaled $\bar{x}$-$\tau$) frame contains both
``incoming" and ``outgoing" components in the rescaled
$\bar{x}$-$\tau$ (original $x$-$t$) frame.

Instanton method is another technique commonly used in the
calculation of the decay rate of a metastable state \cite{Coleman}.
This semiclassical method amounts essentially to finding the
imaginary part of the ground state energy of the system.  Again,
it is not suitable for our case for the same first reason given
above for the failure of the complex eigenvalue method.

Now that the two most common methods fail to suit our purpose, we
have to look for alternatives.  Fortunately, a different method
exists, namely, the scattering state method (or virtual level
method, as Fermi called it) \cite{Fermi}. This method is much less
well known and seldom used in the literature \cite{Holstein2}.
However, it is conceptually the most satisfying one of all the
methods.  In this method, one first constructs the initial
confining state, which is not viewed as an eigenstate, as a linear
superposition of scattering states with real energies, and follows
its evolution in time. In the course of this evolution, no energy
will become complex.  Unlike the Gamow states, the scattering
states contain both incoming and outgoing components in the region
into which the particle escapes. It is this feature of the method
that makes it most suitable for our present problem. The method is
easily adapted to Eq.(\ref{eqs-7}) by taking the scattering states
as the states (\ref{eqs-8}) with real values of $\bar E$.

In the following two sections, we apply the scattering state
method to two simple examples of the class of scaling potentials
discussed in this section.  As the scattering state method is not
so well known in the literature, we think it appropriate to give
some details in order to make our paper self-contained.  We shall
follows the procedures given in \cite{Holstein} which are slightly
adapted to our needs.

\medskip

\section{Moving delta-function potential}

Our first example is a uniformly moving delta-function potential
\begin{eqnarray}
 V(x,t) = \left\{
  \begin{array}{ll}
    \infty & \ , \quad x \leq 0 ~;\\
    \frac{\bar{V}_{0}}{L(t)}\delta (x-a(t))
    = \frac{\bar{V}_{0}}{L^2(t)}\delta (\frac{x}{L(t)}-\bar{a})
    & \ ,\quad x > 0~,
  \end{array} \right.
\end{eqnarray}
where $a(t)=\bar{a}L(t)>0$ gives the location of the
delta-potential. This is of the class of potentials defined in the
last section and corresponds in the rescaled frame to
$\bar{V}(\bar{x})=\infty$ for $\bar{x}<0$ and
$\bar{V}(\bar{x})=\bar{V}_0\delta (\bar{x}-\bar{a})$. We must have
$\Phi (\bar{x})=0$ in the region $\bar{x}<0$.  In the region
$\bar{x}>0$, Eq.(\ref{eqs-9}) is
\begin{eqnarray}
 -\frac{\hbar^2}{2m}\frac{d^2\Phi(\bar{x})}{d\bar{x}^2}
 + \bar{V}_{0} \delta(\bar{x}-\bar{a}) \Phi(\bar{x})
 = \bar{E}\Phi(\bar{x})~.
\label{eqs-13}
\end{eqnarray}
Its general solutions are
\begin{eqnarray}
 \Phi(\bar{x}) = \left\{
  \begin{array}{ll}
    \sin(\bar{k}\bar{x})~, &\quad 0 < \bar{x} < \bar{a}~; \\
    C\cos(\bar{k}\bar{x}+\theta)~,  &\quad  \bar{a} < \bar{x}~,
  \end{array} \right.
\end{eqnarray}
where $\bar{k} =\sqrt{2m\bar{E}}/\hbar$, $C$ is a real constant,
and $\theta$ a phase angle. Note that we have chosen the wave
function to be real, and included incoming wave component in the
region $\bar{x}>\bar{a}$. This ensures that $\bar k$, and hence
$\bar E$, is always real. The wave function and its first
derivative satisfy the following boundary conditions at
$\bar{x}=\bar{a}$:
\begin{eqnarray}
 \Phi(\bar{x}=\bar{a}^{+}) = \Phi(\bar{x}=\bar{a}^{-})~,
\label{eqs-15}
\end{eqnarray}
and
\begin{eqnarray}
 \left. \frac{d\Phi(\bar{x})}{d\bar{x}} \right|_{\bar{x}=\bar{a}^{+}}
 - \left. \frac{d\Phi(\bar{x})}{d\bar{x}} \right|_{\bar{x}=\bar{a}^{-}}
 = \frac{2m}{\hbar^2}\bar{V}_{0}\Phi(\bar{x}=\bar{a})~.
\label{eqs-16}
\end{eqnarray}
>From these relations the coefficient $C$ can be determined as a
function of $\bar{k}$,
\begin{eqnarray}
 C^2(\bar{k})
 = \sin^2(\bar{k}\bar{a}) + \left( \cos(\bar{k}\bar{a})
 + \frac{2m\bar{V}_{0}}{\hbar^2\bar{k}\bar{a}}\sin(\bar{k}\bar{a})
 \right)^2~.
\label{eqs-17}
\end{eqnarray}
Physically, the value of $C^2(\bar{k})$ can be interpreted as the
ratio of the probability of finding particles in the region
$\bar{x}>\bar{a}$ to the probability within the confined region
$0<\bar{x}<\bar{a}$ for a particular $\bar{k}$.  The general shape
of $C^2 (\bar{k})$ is shown in Fig.~1, from which we can assert
that the particle can be  trapped within the confined region only
when $C^2$ assumes one of its minima, which occur only in the
neighborhood of some specific values of $\bar{k}$.   In these
regions the values of $C^2(\bar{k})$ are extremely small. From
Eq.(\ref{eqs-17}) it is obvious that these minima will center
around $\bar{k}_{n}=n\pi/\bar{a}$ ($n=1,2,\ldots$) (i.e. $
\sin(\bar{k}_{n}\bar{a}) = 0$) as long as $\bar{V}_0$ is large
enough so that $2m\bar{V}_0/\hbar^2\gg n\pi$.

Below we shall restrict our discussions to the case of large
$\bar{V}_0$. In this case approximate analytic expressions can be
obtained and compared with the corresponding results in the
time-independent case \cite{Nussen}.  According to the scattering
state method, one constructs confining states in the potential
well by taking suitable superposition of the scattering states
with $\bar{k}$ in the neighborhood of $\bar{k}_n$. To this end,
let us first expand $C^2$ about
$\bar{E}_{n}=\hbar^2\bar{k}_n^2/2m$ (we revert to the variable
$\bar{E}$ below)
\begin{eqnarray}
 C^2(\bar{E}) &\approx&
 \left(\frac{m\bar{a}}{\hbar^2\bar{k}_{n}}\right)^2
 \left[ 1+\left(\frac{2m\bar{V}_{0}}{\hbar^2\bar{k}_{n}}\right)^2\right]
 \left( \bar{E}-\bar{E}_{n}+ \delta \right)^2 +
 \left[1+\left(\frac{2m\bar{V}_{0}}{\hbar^2\bar{k}_{n}}\right)^2\right]^{-1}
 \\ \nonumber \\
 &\equiv& G^2\left( \Delta + \delta \right)^2 + F^2~,
\label{eqs-20}
\end{eqnarray}
where $\Delta = \bar{E}-\bar{E}_{n}$ and the constants
\begin{eqnarray}
 \delta &=& \frac{2\bar{V}_{0}}{\bar{a}}
 \left[
 1+\left(\frac{2m\bar{V}_{0}}{\hbar^2\bar{k}_{n}}\right)^2\right]^{-1}\ ,
 \bigskip \\
 G^2 &=& \left(\frac{m\bar{a}}{\hbar^2\bar{k}_{n}}\right)^2
 \left[ 1+\left(\frac{2m\bar{V}_{0}}{\hbar^2\bar{k}_{n}}\right)^2\right]\ ,
\label{eqs-22}
\bigskip \\
 F^2 &=&
 \left[1+\left(\frac{2m\bar{V}_{0}}{\hbar^2\bar{k}_{n}}\right)^2\right]^{-1}\ .
\label{eqs-23}
\end{eqnarray}
The scattering states with $\bar E$ in the neighborhood of
$\bar{E}_n$ can then be written as
\begin{eqnarray}
\psi_{\Delta}(\bar{x}) = \left\{
 \begin{array}{ll}
    \sqrt{\frac{2}{R}}
    \frac{1}{\sqrt{G^2\left( \Delta + \delta \right)^2 + F^2}}
    \sin(\bar{k}\bar{x})~, &\quad 0 < \bar{x} < \bar{a}~; \medskip \\
    \sqrt{\frac{ 2}{R}} \cos(\bar{k}\bar{x}+\theta)~,  &
 \quad \bar{a} < \bar{x}~.
\end{array} \right.
\label{eqs-24}
\end{eqnarray}
We have quantized our system in the interval $[0,R]$, where $R\gg
\bar a$ which at the end of the calculation will be set to
infinity.  With these scattering states we now construct an
initial state which is completely confined within the well by
taking linear combination of the scattering states with different
$\Delta$ but the same value of $\bar{E}_{n}$,
\begin{eqnarray}
 \Phi(\bar{x},\tau=0) = \sum_{\Delta}c_{\Delta}\psi_{\Delta}(\bar{x})
 = \left\{
  \begin{array}{ll}
    \phi_{n}(\bar{x})~, &\quad \bar{x}< \bar{a}~; \bigskip \\
    0~, &\quad \bar{x}> \bar{a}~.
 \end{array}
\label{eqs-25}
\right.
\end{eqnarray}
The coefficient $c_{\Delta}$ can be calculated from orthogonality of
the states $\psi_{\Delta}(\bar{x})$,
\begin{eqnarray}
 c_{\Delta} = \int_{0}^{R}d\bar{x} \psi_{\Delta}(\bar{x})\Phi(\bar{x},0)
 = \sqrt{\frac{2}{R}}\frac{1}{\sqrt{G^2(\Delta +\delta)^2 +F^2}}
 \int_{0}^{\bar{a}}d\bar{x}\ \sin(\bar{k}\bar{x})\phi_{n}(\bar{x})\ .
\end{eqnarray}
Choosing
\begin{eqnarray}
 \phi_{n}(\bar{x}) \approx \sqrt{\frac{2}{\bar{a}}}
 \sin \left( \frac{n\pi \bar{x}}{\bar{a}}\right)
 \ \ ; \ \ n=1,2,3,\ldots
\end{eqnarray}
we get
\begin{eqnarray}
 c_{\Delta}
 &\approx& \frac{1}{\sqrt{R\bar{a}}}
 \frac{2}{\sqrt{G^2(\Delta+\delta)^2 +F^2}}
 \int_{0}^{\bar{a}}d\bar{x}\ \sin(\bar{k}\bar{x})
 \sin \left( \frac{n\pi \bar{x}}{\bar{a}}\right)
 \\ \nonumber \\
 &\approx& \sqrt{\frac{\bar{a}}{R}}\frac{1}{\sqrt{G^2(\Delta+\delta)^2 +F^2}}\ .
\end{eqnarray}
The initial state is then given by
\begin{eqnarray}
 \Phi(\bar{x},\tau=0) \approx \sqrt{\frac{\bar{a}}{R}}\sum_{\Delta}
 \frac{1}{\sqrt{G^2(\Delta+\delta)^2 +F^2}}\ \psi_{\Delta}(\bar{x})~.
\end{eqnarray}
>From Eq.(\ref{eqs-10}), (\ref{comb1}) and (\ref{comb2}), the
solution at a later time $\tau$ is
\begin{eqnarray}
 \Phi(\bar{x},\tau) \approx
 \sqrt{\frac{\bar{a}}{R}}\sum_{\Delta}
 \frac{1}{\sqrt{G^2(\Delta+\delta)^2 +F^2}}\ \psi_{\Delta}(\bar{x})
 e^{-\frac{i}{\hbar}(\bar{E}_{n}+\Delta)\tau}\ .
\label{eqs-31}
\end{eqnarray}
As the system is quantized in the interval $[ 0,R ]$, we have
$\bar{k}R =n^\prime\pi/2$ where $n^{\prime}$ is a very large
integer ($n^\prime\gg n$). After replacing the sum by an integral
\begin{eqnarray}
 \sum_{\Delta} \ \longrightarrow \
 \int d\Delta \ \frac{R}{\pi\hbar}\sqrt{\frac{2m}{\bar{E}_{n}}}\ ,
\end{eqnarray}
Eq.(\ref{eqs-31}) becomes
\begin{eqnarray}
 \Phi(\bar{x},\tau) \approx
 \frac{R}{\pi\hbar}\sqrt{\frac{2m}{\bar{E}_{n}}}
 \sqrt{\frac{\bar{a}}{R}}
 \int_{-\infty}^{\infty}d\Delta\
 \frac{1}{\sqrt{G^2(\Delta+\delta)^2 +F^2}}\ \psi_{\Delta}(\bar{x})
 e^{-\frac{i}{\hbar}(\bar{E}_{n}+\Delta)\tau}\ .
\label{eqs-33}
\end{eqnarray}
Substituting $\psi_{\Delta}(\bar{x})$ from
Eq.(\ref{eqs-24}), we obtain the approximate wave function of the state
confined in the well ($0 < \bar{x} < \bar{a}$) as
\begin{eqnarray}
\Phi(\bar{x},\tau) &\approx&
\frac{2}{\pi\hbar}\sqrt{\frac{m\bar{a}}{\bar{E}_{n}}}
\sin(\bar{k}_n\bar{x}) e^{-\frac{i}{\hbar}\bar{E}_{n}\tau}
\int_{-\infty}^{\infty}d\Delta\ \frac{e^{-\frac{i}{\hbar}\Delta
\tau}} {G^2(\Delta+\delta)^2 +F^2}
 \nonumber\\
 &=& \frac{2}{\hbar}\sqrt{\frac{m\bar{a}}{\bar{E}_{n}}}
 \frac{\sin(\bar{k}_n\bar{x})}{|FG|}
 e^{-\frac{i}{\hbar}\bar{E}_{n}\tau}\
 e^{-\frac{1}{\hbar}|\frac{F}{G}|\tau}\ .
\label{Phi}
\end{eqnarray}

For metastable system an important quantity is the non-decay
probability $P(t)$ that the particle is still in the well at time
$t$ if it is initially confined in the well at $t=0$ ($P(t=0)=1$).
In our case $P(t)$ is defined as
\begin{eqnarray}
P(t)&=&\frac{\int_0^{a(t)} |\Psi (x,t)|^2 dx} {\int_0^{a(0)} |\Psi
(x,0)|^2  dx}\nonumber\\ &=&\frac{\int_0^{\bar a} |\Phi
(\bar{x},\tau)|^2 d\bar{x}}{\int_0^{\bar a} |\Phi (\bar{x},0)|^2
d\bar{x}}~. \label{P}
\end{eqnarray}
>From Eq.(\ref{Phi}) we find
\begin{eqnarray}
P(t)\sim \exp(-\gamma_n(t))~,
\end{eqnarray}
where
\begin{eqnarray}
\gamma_n(t) &=& 2\left| \frac{F}{G}\right| \frac{t}{L_{0}L(t)}
\bigskip \\
&=& 2 \left( \frac{\hbar^2\bar{k}_{n}}{m\bar{a}}\right)
\left[ 1+ \left(
\frac{2m\bar{V}_{0}}{\hbar^2\bar{k}_{n}}\right)^2\right]^{-1}
\frac{t}{L_{0}L(t)}~.
\label{eqs-38}
\end{eqnarray}
We have used Eq.(\ref{eqs-6}), (\ref{eqs-22}) and (\ref{eqs-23})
to obtain the result (\ref{eqs-38}).  We note that for expanding
potential ($v>0$)
\begin{eqnarray}
\gamma_n(t) \to 2\left| \frac{F}{G}\right| \frac{1}{L_{0}v}
\end{eqnarray}
as $t\to \infty$.  Unlike the stationary case ($v=0$), there is a
small but finite probability that the particle does not tunnel out
of the well.  This result is reasonable, since as the barrier
moves away from $x=0$ it leaves more room for the particle to stay
within the well.

For $\bar{V}_{0}$ much
larger than the characteristic value of $\bar{k}_n$ of the escaping particles,
$\gamma(t)$ becomes
\begin{eqnarray}
\gamma_n(t) \ \longrightarrow\ \left[
 \frac{(\hbar^2\bar{k}_{n})^3}{2m^3\bar{a}\bar{V}_{0}^2}\right]
 \frac{t}{L_{0}L(t)}
 = \left[
 \frac{\hbar^6(n\pi)^3}{2m^3\bar{a}^4\bar{V}_{0}^2}\right]
 \frac{t}{L_{0}L(t)}~,
\label{eqs-39}
\end{eqnarray}
where $\bar{k}_{n}=n\pi/\bar{a}$ has been substituted. It is
proper to compare Eq. (\ref{eqs-39}) with the corresponding result
in the stationary case ($v=0$).  In the limit $v\to 0$, we have
$L(t)\rightarrow L_{0}$, $\bar{a}\rightarrow a/L_{0}$,
$\bar{E}_{n}\to E_{n}L_0^2$, and $\bar{k}_{n}\rightarrow
k_{n}L_{0}$, where $E_n$ is the corresponding energy in the static
frame, and $k_n=\sqrt{2mE_{n}}/\hbar$.  In this limit $\gamma
(v=0)$ is directly proportional to the time $t$.  We can therefore
define a decay rate by $\Gamma_n\equiv \gamma_n (v=0)/t$, which in
this case is
\begin{eqnarray}
 \Gamma_n =
 \frac{\hbar^{6}k_{n}^3}{2m^{3}a(\frac{\bar{V}_{0}}{L_{0}})^2}
 = \frac{\hbar^6(n\pi)^3}{2m^{3}a^{4}(\frac{\bar{V}_{0}}{L_{0}})^2}\ .
\label{eqs-40}
\end{eqnarray}
Eq.(\ref{eqs-40}) is consistent with the result obtained by the complex
eigenvalue method in \cite{Nussen} for a static delta-potential located at
$x=a$ with strength $\bar{V}_0/L_0$.

\section{Moving square barrier potential}

Next, we consider a moving barrier potential
\begin{eqnarray}
  V(x,t) = \left\{
  \begin{array}{ll}
    \infty~, & \quad x \leq 0~; \\
    \frac{1}{L^2(t)}\bar{V}_{0}~,
    & \quad a(t) < x < b(t)~;
    \\
    0~, & \quad x>b(t)~,
  \end{array} \right.
\end{eqnarray}
with $a(t)=\bar{a}L(t)$ and $b(t)=\bar{b}L(t)$ ($\bar{a}$ and $\bar{b}$
are two positive constants).
When the problem is transformed to the rescaled frame, it is
equivalent to solving Eq.(\ref{eqs-9}) with a stationary potential
\begin{eqnarray}
  \bar{V}(\bar{x}) = \left\{
  \begin{array}{ll}
    \infty~, & \quad \bar{x} \leq 0~; \\
    \bar{V}_{0}~,
    & \quad \bar{a} < \bar{x} < \bar{b}~;
    \\
    0~, & \quad \bar{x}>\bar{b}~.
  \end{array} \right.
\end{eqnarray}
The general solutions are
\begin{eqnarray}
 \Phi(\bar{x}) = \left\{
  \begin{array}{lll}
    \sin(\bar{k}\bar{x})~, & \quad 0 < \bar{x} < \bar{a}~,
    & ~\bar{k}=\sqrt{\frac{2m\bar{E}}{\hbar^2}}~; \\
    A e^{\bar{k}'\bar{x}} + B e^{-\bar{k}'\bar{x}}~, &
    \quad \bar{a} < \bar{x} < \bar{b}~,
    & ~\bar{k}'=\sqrt{\frac{2m(\bar{V}_{0}-\bar{E})}{\hbar^2}}~;\\
    C\cos(\bar{k}\bar{x}+\theta)~,  & \quad\bar{a} < \bar{x} ~.&
  \end{array} \right.
\label{eqs-43}
\end{eqnarray}
Here $A,B$ and $C$ are real constants, and $\theta$ a phase angle.
As before, we have set the solutions real in the whole region to
ensure that $\bar{k}$ is always real. We also require the
solutions and their derivatives be continuous at the boundaries
$\bar{x}=\bar{a}$ and $\bar{b}$.  These boundary conditions
determine the values of the coefficients $A$, $B$ and $C$ as
functions of $\bar{k}$:
\begin{eqnarray}
 A(\bar{k}) &=& \frac{1}{2}e^{-\bar{k}'\bar{a}} \left[
 \sin(\bar{k}\bar{a}) +
 \frac{\bar{k}}{\bar{k}'}\cos(\bar{k}\bar{a}) \right]\ ,
 \label{A} \\
 B(\bar{k}) &=& \frac{1}{2}e^{\bar{k}'\bar{a}}\ \left[
 \sin(\bar{k}\bar{a}) -
 \frac{\bar{k}}{\bar{k}'}\cos(\bar{k}\bar{a}) \right]\ ,
 \label{B} \\
 C^2(\bar{k}) &=& \left(1+\frac{\bar{k}^{\prime 2}}{\bar{k}^2}\right)
 e^{2\bar{k}'\bar{b}}A^2
 + 2\left(1-\frac{\bar{k}^{\prime 2}}{\bar{k}^2}\right) A B
 + \left(1+\frac{\bar{k}^{\prime 2}}{\bar{k}^2}\right)
 e^{-2\bar{k}'\bar{b}}B^2\ .
\label{C}
\end{eqnarray}
The general shapes of $A(\bar{k})$ and $C^2(\bar{k})$ are shown in
the Fig.~2. We see that metastable states of the system will occur
only in a finite number of neighborhood of $\bar{k}_n$
($n=1,2,\ldots$) such that $A(\bar{k}_n)= 0$.   The roots
$\bar{k}_n$ satisfy
\begin{eqnarray}
 \sin(\bar{k}_{n}\bar{a}) +
 \frac{\bar{k}_{n}}{\bar{k}_{n}'}\cos(\bar{k}_{n}\bar{a})
 = 0~.
\label{eqs-47}
\end{eqnarray}
Eq.(\ref{C}) implies that $C^2(\bar{k})$ is minimal at
$\bar{k}_n$. For a given $\bar{V}_{0}$, the number of roots
$\bar{k}_{n}$ is restricted by the condition that $\bar{k}_{n}'$
in Eq.(\ref{eqs-43}),
\begin{eqnarray}
 \bar{k}_{n}'=\sqrt{\frac{2m\bar{V}_{0}}{\hbar^2}-\bar{k}_{n}^2}
\end{eqnarray}
must be real. Hence the possible values of $\bar{k}_n$ can only
lie in the interval $( 0, \sqrt{\frac{2m\bar{V}_{0}}{\hbar^2}})$.
For instance, there are only two roots for the parameters assumed
in Fig.~2.

Let us now expand the coefficients $A(\bar{k})$ and $B(\bar{k})$
about $\bar{E}_{n}\ (=\hbar^2\bar{k}_{n}^2/2m)$
\begin{eqnarray}
 A(\bar{E}) &\approx& \left[ \frac{d A}{d \bar{E}}
 \right]_{\bar{E}=\bar{E}_{n}}
 ( \bar{E}- \bar{E}_{n})\ ,
 \label{eqs-48}
 \bigskip \\
 B(\bar{E}) &\approx& B(\bar{E}_{n})
\label{eqs-49}\ .
\end{eqnarray}
Inserting Eq.(\ref{eqs-48}) and (\ref{eqs-49}) into Eq.(\ref{C}),
and after some tedious calculations we find that $C^2(\bar{E})$
behaves in the neighborhood of $\bar{E}_{n}$ as
\begin{eqnarray}
 C^2(\bar{E}) = G^2(\Delta +\delta)^2 + F^2\ ,
\label{eqs-50}
\end{eqnarray}
where $\Delta = \bar{E}-\bar{E}_{n}$ as in the previous example,
and the constants in the present case are
\begin{eqnarray}
 G^2 &=& \frac{1}{4}\left( \frac{m\bar{a}}{\hbar^2\bar{k}_{n}}\right)^2
 \left( 1+\frac{\bar{k}_{n}^{\prime 2}}{\bar{k}_{n}^2}\right)
 \left[ \cos(\bar{k}_{n}\bar{a}) -
 \frac{\bar{k}_{n}}{\bar{k}_{n}'}\sin(\bar{k}_{n}\bar{a})
 \right]^2
 e^{2\bar{k}_{n}'(\bar{b}-\bar{a})}\ ;
\label{eqs-51}
\bigskip \\
 F^2 &=& \left( 1+\frac{\bar{k}_{n}^{2}}{\bar{k}_{n}^{\prime 2}}\right)^{-1}
 \left[ \sin(\bar{k}_{n}\bar{a}) -
 \frac{\bar{k}_{n}}{\bar{k}_{n}'}\cos(\bar{k}_{n}\bar{a}) \right]^2
 e^{-2\bar{k}_{n}'(\bar{b}-\bar{a})}
\label{eqs-52}
\end{eqnarray}
and
\begin{eqnarray}
 \delta = \left( \frac{\hbar^2\bar{k}_{n}}{m\bar{a}}\right)
 \left( \frac{\bar{k}_{n}^2-\bar{k}_{n}^{\prime 2}}
 {\bar{k}_{n}^2+\bar{k}_{n}^{\prime 2}} \right)
 \left(
 \frac{\bar{k}_{n}'\sin(\bar{k}_{n}\bar{a})-\bar{k}_{n}\cos(\bar{k}_{n}\bar{a})}
 {\bar{k}_{n}'\cos(\bar{k}_{n}\bar{a})-\bar{k}_{n}\sin(\bar{k}_{n}\bar{a})}
 \right)
 e^{-2\bar{k}_{n}'(\bar{b}-\bar{a})}\ .
\end{eqnarray}
\bigskip

Once we obtain the relation (\ref{eqs-50}), we can construct the
scattering states relevant to this metastable system following
exactly the same procedures as those in the previous section. The
non-decay probability $P(t)$ of finding the particle within the
confined region $(0<\bar{x}<\bar{a})$ at time $t$ is again of the
from $P(t)\sim \exp(-\gamma_n(t))$, where $\gamma_n(t)$ is now
given by
\begin{eqnarray}
\gamma_{n}(t) &=& 2\left| \frac{F}{G} \right| \frac{t}{L_{0}L(t)}
\nonumber \\
 &=& \frac{8\hbar^2\bar{k}_{n}^3}{m\bar{a}}
 \left( \frac{\bar{k}_{n}'}{\bar{k}_{n}^2+\bar{k}_{n}^{\prime 2}}
 \right)^2
 e^{-2\bar{k}_{n}'(\bar{b}-\bar{a})}\frac{t}{L_{0}L(t)}
\label{eqs-56}\ ,
\end{eqnarray}
where we have used Eq.(\ref{eqs-51}), (\ref{eqs-52}) and
(\ref{eqs-47}). When the barrier height is much larger than the
characteristic ``energy" of the escaping particles $(\bar{V}_{0}
\gg \bar{E}_{n})$, which is equivalent to the relation
$\bar{k}_{n}' \gg \bar{k}_{n}$, $\gamma_n(t)$ becomes
\begin{eqnarray}
\gamma_{n}(t) \ \longrightarrow \ \frac{8\hbar^2}{m\bar{a}}
\frac{\bar{k}_{n}^3}{\bar{k}_{n}^{\prime 2}}
e^{-2\bar{k}_{n}'(\bar{b}-\bar{a})}\frac{t}{L_{0}L(t)}\ .
\label{eqs-57}
\end{eqnarray}
As in the previous example, for positive $v>0$ (the expanding
case) one finds a small but finite probability that the particle
does not tunnel out of the well at large time . In this case not
only does the barrier leave more room for the particle to stay
within the well as it moves away from the $x=0$, but its width
also become thicker, thus making tunneling difficult.

In order to transform the result (\ref{eqs-57}) to the stationary
one, we again make use of the same substitutions as given at the
end of the last section, with the addition of $\bar{b}\to b/L_0$
and $\bar{k}_n'\to k_n'L_{0}$, where
$k_n'=\sqrt{2m((\bar{V}_0/L_0^2)-E_{n})}/\hbar$.  Once again
$\gamma_n (v=0)$ is directly proportional to the time $t$, in
which case a decay rate can be defined: $\Gamma_n\equiv \gamma_n
(v=0)/t$.  For the present example we have
\begin{eqnarray}
 \Gamma_{n}(v=0) =
 \frac{8\hbar^2}{m a}
 \frac{k_{n}^3}{k_{n}^{\prime 2}}
 \ e^{-2 k_{n}'(b-a)}\ ,
\label{eqs-59}
\end{eqnarray}
which is the same as the result obtained by the complex eigenvalue
method for a square barrier with width $(b-a)$ and height $\bar{V}_0/L_0^2$
\cite{Bohm}.
\medskip

\section{General scaling potentials}

We have calculated the non-decay probabilities of two
non-stationary metastable systems explicitly. The potential
barriers in the rescaled frame considered in these systems assumed
the form of a delta-function and a square barrier. These
calculations can be immediately generalized to barriers with more
general shapes.  Without giving further examples, what we would
like to do here is to discuss briefly a close connection between
the the non-decay probability $P(t)$ of a particle in a metastable
scaling potential $V(x,t)=\bar{V}(x/L(t))/L^2(t)$ and the decay
rate $\Gamma$ of the same particle if it were instead confined in
a static potential well $V(x)=\bar{V}(x)$.

>From the discussions and examples in the previous sections, we
know that the calculations of $P(t)$ is reduced to the
corresponding computations in a static potential $\bar{V}
(\bar{x})$ in the rescaled frame.  Now the later task would be
exactly the same as that carried out in the potential
$V(x)=\bar{V}(x)$ in ordinary coordinates.  The only difference,
as seen from the previous two examples, is that all ordinary
parameters, such as $E$, $k$, $k^\prime$, $t$, $a$, etc, are
replaced by the corresponding rescaled ones, i.e. $\bar E$, $\bar
k$, $\bar{k}^\prime$, $\tau$, $\bar a$, etc.  Application of the
scattering state method to the general alpha-decay type of
potential $V(x)$ in normal coordinates has been given in
\cite{Holstein}, and can be carried over directly.  Following
\cite{Holstein} the important step is to determine the discrete
values $E_n$ (or equivalently $k_n$) that minimize the amplitude
$C$ of the wave function in the region outside the well.  Consider
a confining state constructed with $E$ centered around a specific
$E_n$.  Minimization of $C$ then gives the two functions $F(E_n)$
and $G(E_n)$ (other parameters in $F$ and $G$ are not indicated).
The non-decay probability in $V(x)$ is then given by
$\exp(-\Gamma_n t)$, where the decay rate $\Gamma_n$ is
\begin{eqnarray}
\Gamma_n (E_n)= 2\left| \frac{F(E_n)}{G(E_n)}\right|~.
\end{eqnarray}
Suppose all these computations have been done in ordinary
coordinates.  Then one can immediately write down the expression
of the non-decay probability $P(t)\sim \exp(-\gamma_n(t))$ for the
scaling potential $V(x,t)$ as
\begin{eqnarray}
 \gamma_n(t)&=&2\left| \frac{F(\bar{E}_n)}{G(\bar{E}_n)}\right|
 \frac{t}{L_{0}L(t)}\nonumber\\
 &=&\Gamma_n (\bar{E}_n)\frac{t}{L_{0}L(t)}~.
 \label{rate}
 \end{eqnarray}
 Here the functional form of the decay rate $\Gamma_n$ is taken
 over directly, but with all the parameters replaced by the
 corresponding rescaled ones.
 Eq.(\ref{rate}) gives the connection between the non-decay
 probability in $V(x,t)=\bar{V}(x/L(t))/L^2(t)$
 and the decay rate in $V(x)=\bar{V}(x)$.
 Finally, we note here that, in the non-moving limit $v=0$,
 $V(x,t)$ becomes $V(x)=\bar{V}(x/L_0)/L^2_0$.  Setting $v=0$ in Eq.(\ref{rate})
 then gives the decay rate in this potential: $\Gamma_n
 (E_n)/L^2_0$, as we had seen in the previous cases.

\section{Conclusion}

In this paper we consider quantum metastability in a class of
moving potentials introduced by Berry and Klein. Potential in this
class has its height and width scaled in a specific way so that it
can be transformed into a stationary one.  In deriving the
non-decay probability of the system, we employed a method which is
less well known but conceptually more satisfactory, namely, the
method of scattering states. Non-decay probabilities in a moving
delta-potential and a moving square barrier potential were
derived.  We also give a connection between the non-decay
probability in a general scaling potential and the decay rate in a
related static potential.
\bigskip

\begin{acknowledgments}
This work was supported in part by the Republic of China through
Grant No. NSC 92-2112-M-032-005.
\end{acknowledgments}

\newpage
\centerline{\bf Figure Captions}
\begin{description}
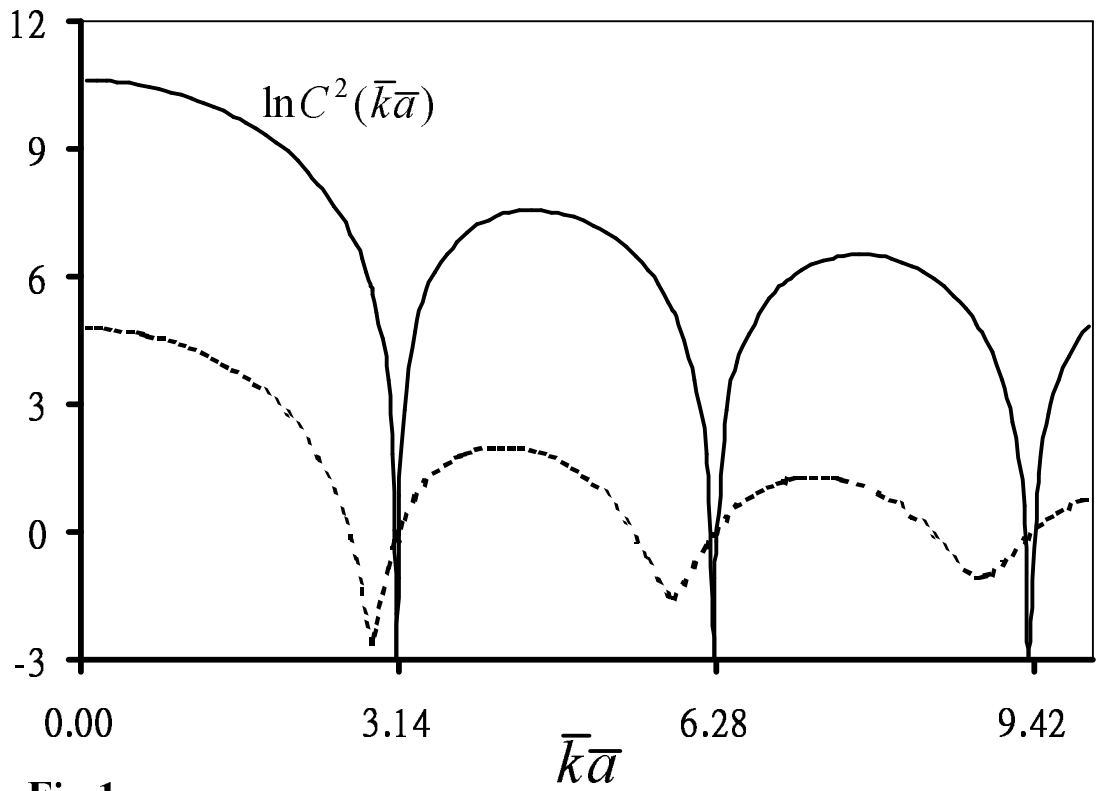
\item[Figure 1.] The shape of $\ln C^{2}(\bar{k}\bar{a})$ in
Eq.(\ref{eqs-17}) as a function of $\bar{k}\bar{a}$ for
$2m\bar{V}_{0}/\hbar^2=10$ (dotted line) and $200$ (solid line).
The minima of $\ln C^{2}(\bar{k}\bar{a})$ will center around
$\bar{k}_{n}\bar{a} = n\pi$ ($n=1,2,\ldots$) for large values of
$2m\bar{V}_{0}/\hbar^2$.

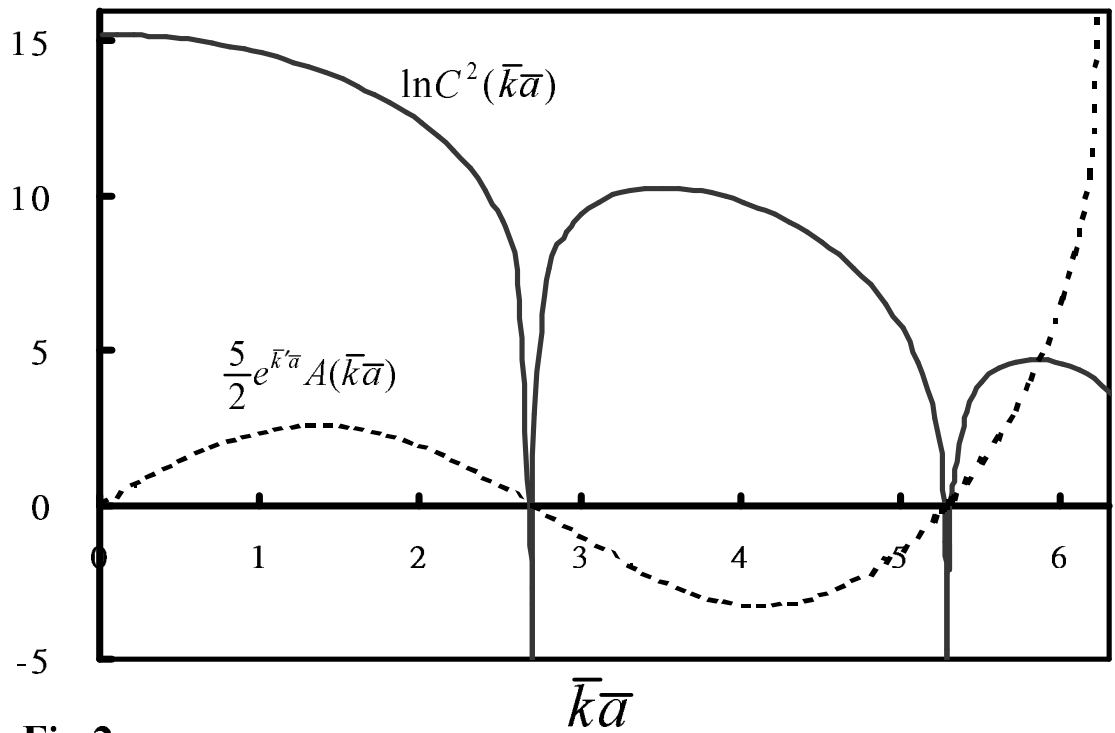
\item[Figure 2.]  The shapes of $\ln
C^{2}(\bar{k}\bar{a})$ (solid line) in Eq.(\ref{C}) and
$5e^{\bar{k}'\bar{a}}A(\bar{k}\bar{a})/2$ (dotted line) in
Eq.(\ref{A}) as function of $\bar{k}\bar{a}$ for
$\bar{b}=2\bar{a}$ and $2m\bar{V}_{0}\bar{a}^2/\hbar^2=40$ . It
shows that the minima of $\ln C^{2}(\bar{k}\bar{a})$ only occur
in a finite number of neighborhood of $\bar{k}_{n}\ (n=1,2,\ldots)$
such that $A(\bar{k}_{n}\bar{a}) =0$.
\end{description}
\end{document}